\documentclass[doublecol,linenumbers]{epl2} 

\newcommand {\beq}{\begin{eqnarray}}
\newcommand {\eeq}{\end{eqnarray}}

\newcommand \sk{\sum_{\mathbf{k}}}
\newcommand \ek{\epsilon_{\mathbf{k}}}
\newcommand \mk{\mathbf{k}}

\title{Coleman-Weinberg mechanism  in
spinor Bose-Einstein \\ condensates}

\author{Shun Uchino}

\institute{                    
  DPMC-MaNEP, University of Geneva,
24 Quai Ernest-Ansermet, CH-1211 Geneva, Switzerland
}

\abstract{
It is argued that 
a continuous quantum phase transition between different ordered phases
in spinor Bose-Einstein condensates
predicted by the mean-field theory 
is vulnerable to quantum fluctuations. 
By analyzing  Lee-Huang-Yang corrections in the condensate,
we demonstrate that the so-called Coleman-Weinberg mechanism takes place in
such a transition,
that is, the transition becomes the first order by quantum fluctuations.
A jump to be expected in this first-order transition is 
induced by a correction from density fluctuations 
despite a transition between different magnetic properties
with keeping condensation.
We exemplify this with an experimentally relevant case
and show that a measurement of a condensate depletion can be utilized 
to confirm the first-order  transition.}

\begin{document}

\maketitle

{\it Introduction} --- 
A competition among different degrees of freedom changes
the nature of phase transitions qualitatively.
This has been first recognized 
in the elementary particle physics and is nowadays called
Coleman-Weinberg mechanism
\cite{PhysRevD.7.1888} where
the massless scalar field
gains a mass due to  
quantum fluctuations on the gauge field.
In the condensed matter physics,
the similar phenomenon has been rediscovered and is called
fluctuation-induced first-order transitions 
where the thermal fluctuations on the gauge field 
and the director
induce first-order transitions in the type-I superconductor
and in the smectic-A liquid crystal, respectively
\cite{PhysRevLett.32.292}, while a mean-field theory incorrectly predicts 
continuous phase transitions in both systems.

Recently, ultracold atomic gases have provided 
an ideal arena to explore classical and quantum phase transitions.
If we focus attention on a quantum phase transition being a particularly
interesting class \cite{sachdev2007quantum},
superfluid-Mott insulator transition \cite{greiner2002quantum} 
and paramagnet-antiferromagnet transition in
the quantum Ising model \cite{simon2011quantum} have already 
been realized.
In addition to the realizations of systems analogous
to condensed matter physics,
quantum phase transitions
in new states of matters, which have yet to be discussed in other
fields, can also be achieved.

Spinor Bose gas is one of such systems and has been studied because of  
richness of realized phases and a variety of
quantum phase transitions among different ordered phases 
\cite{kawaguchi2012spinor,ueda2012bose,RevModPhys.85.1191}.
For example, in  the system on a lattice \cite{
PhysRevLett.88.163001,PhysRevLett.93.120405,PhysRevA.70.063610,
PhysRevLett.94.110403,
PhysRevLett.95.240404,PhysRevA.74.035601,
PhysRevB.77.014503,PhysRevLett.102.140402}, 
it has been shown that contrary to the spinless case,
the superfluid-Mott insulator transition is first-order
or continuous depending on the spin-dependent coupling and
filling \cite{PhysRevA.70.063610,
PhysRevLett.94.110403, PhysRevB.77.014503,
PhysRevLett.102.140402}. 
 Even in the case with weak couplings where
a Bose-Einstein condensate (BEC) is always present
and the theory of weakly-interacting BECs
can be applied \cite{PhysRevLett.81.742,JPSJ.67.1822},
interesting properties come out.
One of the nontrivial predictions at the mean-field level 
is emergence of
continuous quantum phase transitions 
between different ordered phases 
while most of the transitions
are first order \cite{stenger1998spin,kawaguchi2012spinor,RevModPhys.85.1191}.
Seemingly, this result is counter-intuitive
since it is thought that
a phase transition between different ordered phases is first order
\footnote{
An exception is deconfinement criticality discussed in
quantum spin systems. See e.g., T. Senthil, L. Balents, S. Sachdev,
A. Vishwanath, and M. P. A. Fisher, J. Phys. Soc. Jpn., \textbf{74},
1, (2005). }
due to the observation that 
in many cases, order parameters at the phase
boundary change abruptly  
\cite{Goldenfeld.1992,sachdev2007quantum}. 
At the same time, 
there exist 
situations that order parameter in each phase
is smoothly connected at a boundary, which
permits  
a continuous transition at the mean-field level
and has motivated studies of quench dynamics
\cite{PhysRevLett.98.160404,PhysRevLett.99.120407,PhysRevLett.99.130402,PhysRevA.76.043613,damski2009quantum,0953-8984-25-40-404212}
 \textit{ \`a la}
Kibble and Zurek \cite{kibble1976topology,zurek1985cosmological}.

In this Letter, we show that the Coleman-Weinberg mechanism takes place
in an experimentally verifiable spin-1 BEC.
It is demonstrated that the continuous transition predicted at the mean
field becomes first order, which is driven by a correction from density
fluctuations 
while a BEC is maintained during the transition in which 
a change of magnetic properties is concerned.
In this transition, the Bogoliubov modes from the transverse spin sector
and charge sector play roles similar to the scalar and gauge fields
in the model by Coleman and Weinberg \cite{PhysRevD.7.1888}, respectively. 
Our finding indicates that the nature of the transition is the
itinerant property of particles and interplay
between charge and spin sectors hidden at the mean-field level.
Since the correction from the density fluctuations 
is dominant in spinor BECs with alkali species,
this first-order phase transition may be experimentally detectable.
We also reveal that a measurement of a condensate depletion is able to
confirm our statement.

{\it Hamiltonian and mean-field phase diagram.}--- 
We consider $N$ spin-$1$ identical bosons with
an $s$-wave scattering in a three-dimensional box \footnote{Quite recently,
a BEC in a box potential has been realized in Refs. 
A. L. Gaunt, T. F. Schmidutz, I. Gotlibovych, R. P.
Smith, and Z. Hadzibabic, Phys. Rev. Lett., \textbf{110}, 200406
(2013);
T. F. Schmidutz, I. Gotlibovych, A. L. Gaunt, R. P.
Smith, N. Navon, and Z. Hadzibabic, Phys. Rev. Lett., \textbf{112},
040403 (2013).}.  
We assume that an external magnetic field is 
applied in the $z$ direction.
Due to global spin conservation along the $z$ axis on time scales of 
experiments, 
the most dominant contribution comes from a
quadratic Zeeman effect 
instead of a linear Zeeman effect
\cite{stenger1998spin,RevModPhys.85.1191}.
Then,
the low-energy Hamiltonian of a spin-$1$ BEC is given by 
\beq
H=\int d\mathbf{x}
(H_{0}+H_{int}),
\eeq
where 
\beq
H_{0}=\phi^{\dagger}_m(\mathbf{x})
\left( -\frac{\hbar^{2}\nabla^2}{2M}+qm^2\right)\phi_m (\mathbf{x}),
\eeq
is the single-particle Hamiltonian with the quadratic Zeeman coupling
$q$, 
and
\beq
H_{int}=\phi^{\dagger}_{m}\phi^{\dagger}_{m'}
(c_0\delta_{mn}\delta_{m'n'}+c_1\mathbf{F}_{mn}\cdot
\mathbf{F}_{m'n'})\phi_{n}\phi_{n'}
\label{eq:int}
\eeq
is the interaction Hamiltonian.
Here,
$\mathbf{F}$ are spin-1 matrices,
and $m$ gives the $z$ component of the hyperfine spin in each atom. 
As seen from Eq. (\ref{eq:int}), 
$c_0$ and $c_1$ are spin-independent and spin-dependent coupling
constants,
respectively.
Thus, $c_0>0$ is necessary for a stable BEC \cite{pitaevskii2003bose}.
Essentially, an alkali species used in spinor BECs 
satisfies this condition along with an additional condition
$c_0\gg|c_1|$ \cite{RevModPhys.85.1191}.
 
In the mean-field theory, 
the field $\phi_{m}$ is treated as $c$-number $\psi_m$
with $\psi_{m}=\sqrt{n}\zeta_m$, where $n$ is the density of the atoms and
variational parameters $\zeta_m$ are assumed to satisfy 
the normalization condition 
$\sum_{m}|\zeta_{m}|^2=1$.
In the case of $c_1<0$, which is relevant to a spin-1 $^{87}$Rb
condensate,
the mean-field solutions are given by 
\cite{stenger1998spin,PhysRevA.75.013607}
\begin{figure}[t]
 \begin{center}
  \includegraphics[width=0.8\linewidth]{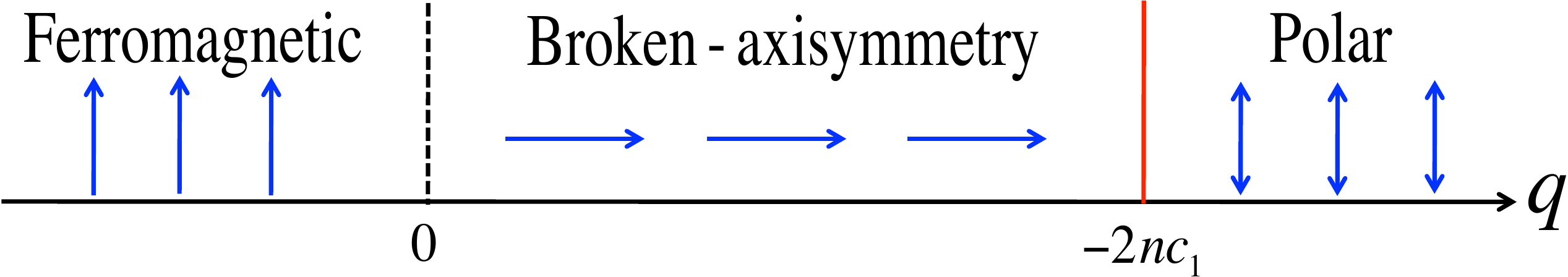}
  \caption{(color online) 
Mean-field zero-temperature
phase diagram with $c_{1}<0$ in the presence of the quadratic
Zeeman effect in spin-1 BEC. Dotted and solid lines at the boundaries
indicate first-order and second-order transitions, respectively.
The solid line (second order) becomes dotted one (first order)
by quantum fluctuations, as demonstrated below.}
  \label{spin-1-phase}
 \end{center}
\end{figure} 
\beq
&&Ferromagnetic \ phase: \ \ 
\mathbf{\zeta}^{F}_m=(1,0,0),\label{ferro}\\
&& Polar \ phase:   \ \ 
\mathbf{\zeta}^{P}_m=(0,1,0),   \label{polar}\\
&&Broken-axisymmetry\ phase: 
\nonumber\\ 
&&\mathbf{\zeta}^{BA}_m=\left(\sqrt{\frac{1}{4}+\frac{q}{8nc_1}},\sqrt{\frac{1}{2}
-\frac{q}{4nc_1}},
\sqrt{\frac{1}{4}+\frac{q}{8nc_1}}\right). \label{ba}
\eeq
The corresponding mean-field phase diagram is shown in 
Fig. \ref{spin-1-phase}.
We point out that each phase may be characterized by
different magnetic orders whose basic properties 
have been confirmed experimentally
\cite{PhysRevLett.92.140403,chang2005coherent,
PhysRevA.84.063625}.
In the ferromagnetic phase,  
the magnetization along the $z$ axis emerges,
while in the broken-axisymmetry phase,
the emergent magnetization is perpendicular to the magnetic field direction.
In relatively large positive $q$, the polar phase emerges where
there is nematicity \cite{zhou2003quantum} but no magnetization.
Here, we focus on the order of phase transitions:
the transition between the ferromagnetic and broken-axisymmetry phases
is first order while
that between the polar and broken-axisymmetry phases
is continuous \cite{stenger1998spin,PhysRevA.75.013607}.
They
can easily be checked by the usual way, namely,
taking derivative of the mean-field ground-state energies
with respect to the couplings \cite{Goldenfeld.1992,PhysRevA.88.043629}.
Intuitively, they can be understood by the fact that
the order parameter abruptly changes at the boundary between
the ferromagnetic and broken-axisymmetry phases
while the smooth change occurs
between the broken-axisymmetry and polar phases.
Therefore, the smooth transition between the broken-axisymmetry and
polar phases is possible while
the abrupt (first-order) transition is expected
in the transition between the ferromagnetic and broken-axisymmetry
phases
due to the level crossing argument \cite{RevModPhys.85.1191}.
In addition, from  the analysis of the lowest order effective theory,
this continuous transition is alleged to lie in the universality class of 
 the 3+1 dimensional
$O(2)$ model due to the absence of the linear Zeeman effect,
which induces the Berry phase term
\cite{PhysRevLett.98.160404,PhysRevB.40.546}.
We note that the model discussed by Coleman and Weinberg 
\cite{PhysRevD.7.1888}
also belongs to this universality class at the mean-field level.

{\it Quantum corrections of ground-state energies.}--- 
Along the lines of the argument by Coleman and Weinberg
\cite{PhysRevD.7.1888,peskin1995introduction},
we consider quantum fluctuation effects at the 1-loop level
in the broken-axisymmetry and polar phases.
In the case of the BEC systems \cite{RevModPhys.76.599}, 
this corresponds to
considering up to the Beliaev \cite{beliaev} or Lee-Huang-Yang (LHY)  
\cite{PhysRev.105.1119,PhysRev.106.1135} theory. 
To this end, we employ the Bogoliubov prescription
\cite{pitaevskii2003bose}
in each phase.
This is achieved by considering the fluctuations of $\phi_{m}$ 
from the $c$-number $\psi_m$
up to the second order at the Hamiltonian level, and
the resultant quantum correction to the ground-state energy 
is nothing but the LHY (1-loop) correction.

The ground-state energy in the polar phase is given by \cite{PhysRevA.81.063632}
\beq
&&E^{P}=E_{MF}^{P}-\frac{1}{2}\sk 
\Bigg[\left( \ek +nc_0- E^{P}_{\mk ,d}\right) 
\nonumber\\ && +2\left(  \ek+q +nc_1 
-E^{P}_{\mk ,f_t}\right) 
-\Big\{
      \frac{(nc_0)^2+2(nc_1)^2}{2\epsilon_{\mk}}\Big\}\Bigg],
\label{eq:gse-p}
\eeq
where $\epsilon_{\mk}=\hbar^2\mk^2/(2M)$, and
\beq
E^{P}_{\mk,d}=\sqrt{\ek (\ek +2nc_0)},
\label{eq:spin-1-p-b1}
\\
E^{P}_{\mk,f_{t}}=\sqrt{(\ek+q) (\ek+q +2nc_1
)},
\label{eq:spin-1-p-b2}
\eeq
are Bogoliubov modes, which originate from density and spin
fluctuations, respectively \cite{PhysRevA.75.013607,PhysRevA.81.063632}.
Here, $E_{MF}^{P}=Nnc_0/2$ with the total particle number $N$ 
is the mean-field energy,
and remaining  terms describe the LHY
correction.
As can be seen from Eq. (\ref{eq:gse-p}),
while each term in the LHY correction has an ultraviolet
divergence,
these divergences are mutually canceled out. Thus 
a finite ground-state energy can be obtained.
This is the consequence of the renormalization.
On the other hand, the ground-state energy in the broken-axisymmetry
phase
is given by \cite{PhysRevA.81.063632}
\beq
E^{BA}=E_{MF}^{BA}
-\frac{1}{2}\sk\Big[
\left(\ek+q/2-E^{BA}_{\mk,f_z}\right)
\nonumber\\ +(2\ek+nc_0-nc_1-E^{BA}_{\mk,d}
-E^{BA}_{\mk,f_{t}})\nonumber\\
-\frac{1}{2\epsilon_{\mk}}\Big\{
\frac{(nc_1-2nc_0)q^2}{2nc_1}+(nc_0+nc_1)^2
\Big\}
\Big],
\label{eq:gse-ba}
\eeq
where $E_{MF}^{BA}=Nn(c_0+c_1+q)/2+Nq^2/(8nc_1)$ is the mean-field energy and
the others are LHY correction in the broken-axisymmetry phase.
The Bogoliubov modes describing density and spin fluctuations 
are \cite{PhysRevA.81.063632}
\beq
E^{BA}_{\mk,d}=\sqrt{\ek^2+n(c_0-c_1)\ek
+2n^2c_1(c_1-c_q)
+E_1(k)},
\label{eq:spin-1-ba-b3}\\
E^{BA}_{\mk,f_{t}}=\sqrt{\ek^2+n(c_0-c_1)\ek
+2n^2c_1(c_1-c_q)
-E_1(k)},
\label{eq:spin-1-ba-b2}\\
E^{BA}_{\mk,f_z}=\sqrt{\ek\left(\ek+q\right)},
\label{eq:spin-1-ba-b1}
\eeq
with $c_q=\frac{q^2}{4n^2c_1}$ and
\beq
E_1(k)=[\{(nc_0+3nc_1)^2-4n^2c_q
(c_0+2c_1)\}\ek^2\nonumber\\
-4n^3c_1(c_0+3c_1)
(c_1-c_q)\ek+\{2n^2c_1(c_1-c_q)
\}^2]^{\frac{1}{2}}.
\eeq
Here,
the transverse spin mode $E^{BA}_{\mk,f_t}$
has a dependence of the spin-independent coupling $c_0$ except at the boundary.
This is in contrast with the polar phase where such a mode depends only
on the spin-dependent couplings, $c_1$ and  $q$. 

{\it Fluctuation-induced first-order transition.}--- 
As far as the mean-field theory is concerned,
the first derivative of the ground-state energy with respect to
the coupling constants has no discontinuity at the boundary 
between the polar and 
broken-axisymmetry phases,
which suggests the continuous transition.
We now consider what happens when 
the quantum fluctuation effect is taken into account.
To see this, we first point out that 
the phase boundary does not move in the presence of the LHY corrections, 
namely, $E^{P}|_{q=-2nc_1}
=E^{BA}|_{q=-2nc_1}$.
This property originates from the fact that the Bogoliubov modes
in the polar phase coincide with those in the broken-axisymmetry phase
at the boundary.

We then consider its derivative with respect to $q$ at the boundary.
The derivative from the polar phase is found to be
\beq
\frac{\partial E^{P}}{\partial q}\Bigg|_{q=-2nc_1}
=\frac{2N\sqrt{M^3}}{3\pi^2\hbar^3}\sqrt{n|c_1|^3},
\eeq
and that from the broken-axisymmetry phase is found to be
\beq
\frac{\partial E^{BA}}{\partial q}\Bigg|_{q=-2nc_1}
=\frac{N\sqrt{M^3}}{3\pi^2\hbar^3}
\left[2\sqrt{n|c_1|^3}+\sqrt{nc_0^3}f(y)\right],
\eeq
where  $y=c_1/c_0$ and
\beq
f(y)=\frac{2+(-y)^{3/2}-2(-y)^{5/2}+4y+3y^2}{(1+y)}.
\label{eq:f-function}
\eeq
Here, to obtain the above, we replaced the summation over $\mk$
in Eqs. (\ref{eq:gse-p}) and (\ref{eq:gse-ba})
by the integral.
Since $f(y)$ 
takes positive values for $c_1<0$ (or equally $y<0$),
there is a jump in the first derivative 
at the boundary.
We note that such a jump is characterized by the LHY correction to
be small considering a weakly-interacting BEC.
This is consistent with the initial assumption, and 
the result should be valid
\footnote{In addition to the smallness of the jump, the high 
rate of condensation and dimensionality are important to apply the
Bogoliubov theory. Since the fluctuations around 4 dimensional $O(2)$
model with the weakly-interacting BEC are concerned, the Bogoliubov
theory can be applied.}
\cite{PhysRevD.7.1888,peskin1995introduction}.

We thus conclude that the fluctuation-induced first-order quantum phase
transition occurs.
While
the transition between the polar and broken-axisymmetry phases
is thought to be magnetic one,
such a jump is proportional to $\sqrt{nc_0^3}$, that is, 
the LHY correction from density fluctuations.  
In the broken-axisymmetry phase, however,
the order parameter (\ref{ba}) is the so-called noninert state
\cite{kawaguchi2012spinor,RevModPhys.85.1191},
which even at the lowest order level
causes an admixture between the spin-independent
and spin-dependent couplings
in the excitation spectra in a complicated way.
As a result,
 the transverse spin mode has the dependence of $c_0$,
which reflects the derivative of the ground-state energy and
eventually
leads to the nontrivial first-order quantum phase transition. 
This result would be reasonable once we recall the fact that 
each particle in a spinor BEC is itinerant, 
which is essentially different from
spin systems and allows  to deviate from the continuous transition.
Finally, we point out that while the transition becomes the first order,
the change in the expectation value of the transverse spin $\langle
\mathbf{F}^{\perp}
\rangle$ is still continuous even at the quantum level.
Although this result can be easily checked by using the Bogoliubov 
theory, it may be interpreted as indicating that
the first-order transition is induced by the density fluctuations,
which does not break the spin rotational symmetry.

{\it Measuring first-order transition.}--- 
In the above, the jump in the first derivative 
is proportional to the LHY correction
from density fluctuations, $\sqrt{nc_0^3}$. 
Since the spin-independent coupling $c_0$ is dominant in a spinor
BEC with alkali atoms, this effect may be detectable.
Here, we relate such a jump to a physical observable.
By using the Hellmann-Feynman theorem
\cite{zphys.85.180,PhysRev.56.340} at finite temperature,
we obtain
\beq
\frac{\partial F}{\partial q}=\frac{1}{Z}Tr\left(
e^{-H/(k_{B}T)}\frac{\partial H}{\partial q}\right)
=\langle(N_1+N_{-1})\rangle_{T},
\eeq
where $Z$ is the partition function and
the above is correct even at $T=0$.
Namely, the first derivative on $q$ corresponds to
the number of particles with hyperfine state $m=\pm1$. 
By adding a magnetic field gradient, which causes 
the Stern-Gerlach separation,
the number of the particles in each hyperfine state
is routinely measured
\cite{RevModPhys.85.1191}, which
can also be utilized to
confirm the first-order phase transition between the polar and 
broken-axisymmetry phases.
At the mean-field level, we expect the smooth change of $N_{1}+N_{-1}$
where it takes nonzero value in the broken-axisymmetry phase
and takes zero in the polar phase and at the boundary.
However, at the quantum level, while such a smooth change 
can be continued in each phase, $N_{1}+N_{-1}$ has the jump
at the boundary due to the quantum fluctuations. 
By taking parameters of spin-1 $^{87}$Rb condensates:
$y=-4.56\times 10^{-3}$,
$c_0/(4\pi\hbar^2a_B/M)=100.86$ with the Bohr radius $a_B$, 
we find that
such a jump is about a few $\%$ of 
the total number of particles with a typical value of the density
in cold atoms
to be of the order of $10^{14}$cm$^{-3}$.

We briefly discuss the thermal fluctuation effects in the first-order
quantum phase transition by means of the finite temperature Bogoliubov theory 
\cite{pitaevskii2003bose}.
The free energy in the polar phase is expressed as
\beq
F^{P}=E^{P}+k_BT\sum_{\mk}[
\ln(1-e^{-E^{P}_{\mk,d}/(k_BT)})\nonumber\\
+2\ln(1-e^{-E^{P}_{\mk,f_t}/(k_BT)})]
\eeq
and that in the broken-axisymmetry phase is done as
\beq
&&F^{BA}=E^{BA}+k_BT\sum_{\mk}[
\ln(1-e^{-E^{BA}_{\mk,d}/(k_BT)})\nonumber\\
&&+\ln(1-e^{-E^{BA}_{\mk,f_t}/(k_BT)})
+\ln(1-e^{-E^{BA}_{\mk,f_z}/(k_BT)})].
\eeq
Although it is difficult to obtain analytic expressions of 
the free energies except for low and high temperatures,
we can numerically 
evaluate it and its derivative.
\begin{figure}[t]
 \begin{center}
  \includegraphics[width=0.8\linewidth]{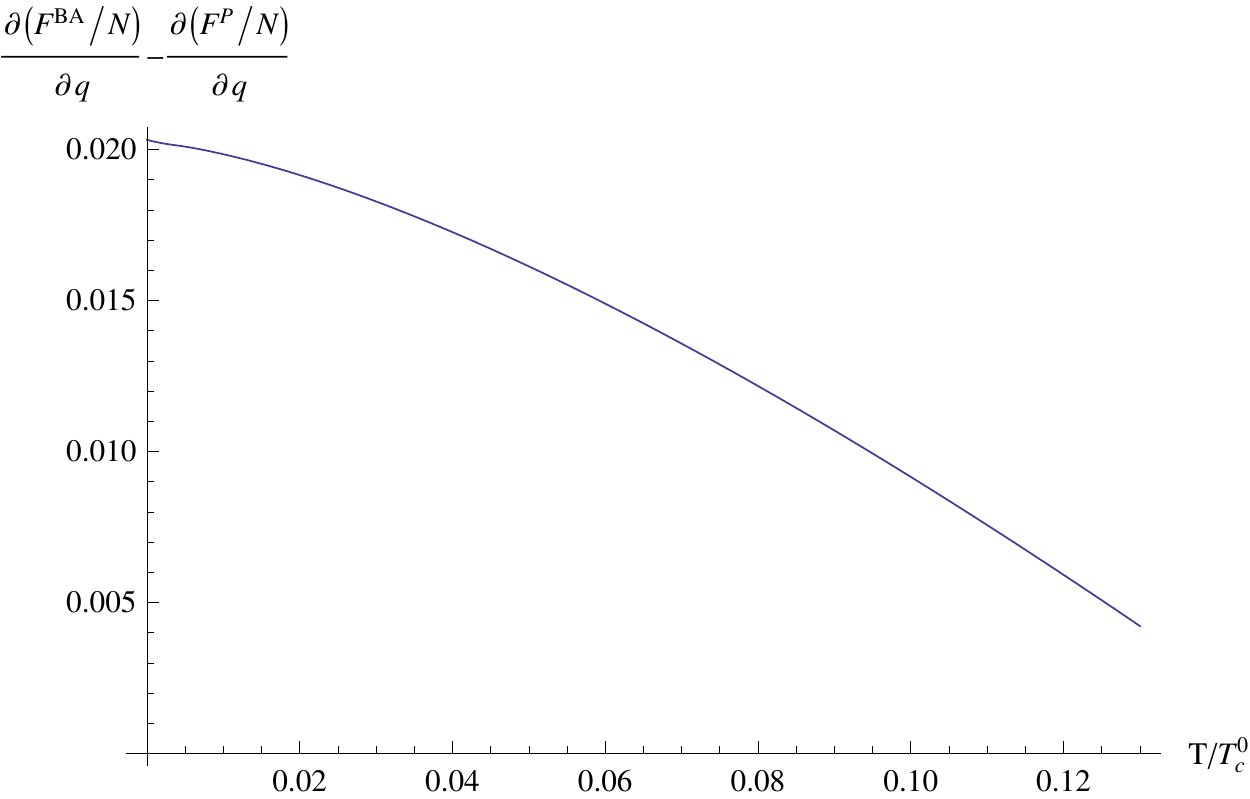}
  \caption{(color online) Typical behavior of the difference of
the first derivatives on the free energies per particle at the boundary 
between the polar and broken-axisymmetry phases
as a function of the temperature. 
Here,
  $T_c^0=\frac{2\pi\hbar^2}{Mk_B}\left(\frac{n}{3\zeta(3/2)}\right)^{2/3}$ 
is the critical temperature of an ideal spinor gas,
and the curve is plotted with the density $n=3\times10^{14}$cm$^{-3}$.}
  \label{finite-t}
 \end{center}
\end{figure} 
As shown in Fig. \ref{finite-t},
the difference of the first derivatives
on the free energies  at the boundary between the polar and
broken-axisymmetry phases gets lower
as the temperature increases, and
the first order phase transition is weakened.
At the same time, if we look at the quantum regime
in the transition
where the finite-temperature Bogoliubov theory is valid,
we can clearly see the first order transition.

We now point out that the jump in the first derivative is enhanced 
on
an optical lattice while such a jump is expected to be small under
the typical current experimental condition in a continuum space.
Unless we consider a too large lattice depth such that
a Mott insulator comes out, the same treatment as we did in the
continuum space can be performed
\cite{PhysRevA.63.053601,rey2003bogoliubov}.
This is due to the fact that the  Bogoliubov
Hamiltonian on the lattice is reduced to that in a continuum space
by substitutions $\epsilon_{\mk}\to 6t-2t\sum_{j=1}^3\cos(k_j)$ and
$c_i\to U_i$ ($i=0,1$)
where $t$ and $U_i$ are the nearest-neighbor hopping matrix
and Hubbard interaction, respectively.
We also note that in addition to the Bogoliubov Hamiltonian, 
the similar form can be obtained for the
condensate depletion on the lattice.
It has been experimentally tested in a spinless BEC
that such an analysis provides a semi-quantitative
description for the condensate depletion
\cite{PhysRevLett.96.180405}.
In the case of spinor BECs, 
the total condensate depletion (sum of the condensate depletion in each
component)
on the lattice has been analyzed in 
\cite{PhysRevA.77.033628}, which numerically estimates that
the total condensate depletion  scales as
$U_0/t$ with a positive exponent and is
of the order of 10$\%$ around the ten recoil energy. 
This result indicates that
 the jump in the first derivative is also expected to be
around 10$\%$ since
the condensate depletion  
in the polar phase takes large value for the $m=0$ component rather than
the $m=\pm1$ components and therefore $N_{1}+N_{-1}$ on the lattice
is still around zero in the polar phase, 
which is in contrast with the case in the
broken-axisymmetry
phase where the large condensate depletion occurs for the $m=\pm1$
components.
Thus, by shifting the magnetic field, we may clearly see the
jump of $N_{1}+N_{-1}$ on the lattice such that
the first-order effect cannot be blurred by the fluctuation of the
magnetic field in experiments.

In conclusion,
we have examined fluctuation effects of 
a would-be continuous quantum phase transition
between polar and broken-axisymmetry phases in a spinor BEC.
We have shown that an interplay between the spin and charge sectors
hidden at the mean-field level leads to the first-order transition
by the Coleman-Weinberg mechanism.
This first-order transition is  characterized by
a  correction from density fluctuations, which 
may pave the way to measure our finding.
We have pointed out that a condensate depletion can be used to
confirm the first-order transition.
Finally, we comment on the other transitions in spinor BECs, 
which are supposed to
be continuous at the mean-field level
\cite{kawaguchi2012spinor,RevModPhys.85.1191}. 
Essential features of such continuum transitions are same as
the transition between the polar and broken-axisymmetry phases: 
the ground-state energies and Bogoliubov modes
coincide at the phase boundaries. 
At the same time, since the forms of the Bogoliubov modes
are different in each phase except for the boundaries,
we can naturally expect that the first derivative of the Bogoliubov modes
on the coupling constants should be different even at the boundaries.
Therefore, such continuum transitions may also become first order.

\acknowledgments
The author thanks T. Giamarchi, M. Nitta, D. Takahashi,
A. Tokuno, and M. Ueda
for fruitful conversations. 
This work is supported by 
the Swiss National Science Foundation under MaNEP and division II.


\begin{thebibliography}{10}
\expandafter\ifx\csname url\endcsname\relax\def\url#1{\texttt{#1}}\fi

\bibitem{PhysRevD.7.1888}
\Name{Coleman S. \and Weinberg E.} \REVIEW{Phys. Rev. D}{7}{1973}{1888}.

\bibitem{PhysRevLett.32.292}
\Name{Halperin B.~I., Lubensky T.~C. \and Ma S.-k.} \REVIEW{Phys. Rev.
  Lett.}{32}{1974}{292}.

\bibitem{sachdev2007quantum}
\Name{Sachdev S.} \Book{Quantum phase transitions} (Cambridge University Press)
  2007.

\bibitem{greiner2002quantum}
\Name{Greiner M., Mandel O., Esslinger T., H{\"a}nsch T.~W. \and Bloch I.}
  \REVIEW{Nature}{415}{2002}{39}.

\bibitem{simon2011quantum}
\Name{Simon J., Bakr W.~S., Ma R., Tai M.~E., Preiss P.~M. \and Greiner M.}
  \REVIEW{Nature}{472}{2011}{307}.

\bibitem{kawaguchi2012spinor}
\Name{Kawaguchi Y. \and Ueda M.} \REVIEW{Physics Reports}{520}{2012}{253}.

\bibitem{ueda2012bose}
\Name{Ueda M.} \REVIEW{Annu. Rev. Condens. Matter Phys.}{3}{2012}{263}.

\bibitem{RevModPhys.85.1191}
\Name{Stamper-Kurn D.~M. \and Ueda M.} \REVIEW{Rev. Mod.
  Phys.}{85}{2013}{1191}.

\bibitem{PhysRevLett.88.163001}
\Name{Demler E. \and Zhou F.} \REVIEW{Phys. Rev. Lett.}{88}{2002}{163001}.

\bibitem{PhysRevLett.93.120405}
\Name{Imambekov A., Lukin M. \and Demler E.} \REVIEW{Phys. Rev.
  Lett.}{93}{2004}{120405}.

\bibitem{PhysRevA.70.063610}
\Name{Krutitsky K.~V. \and Graham R.} \REVIEW{Phys. Rev. A}{70}{2004}{063610}.

\bibitem{PhysRevLett.94.110403}
\Name{Kimura T., Tsuchiya S. \and Kurihara S.} \REVIEW{Phys. Rev.
  Lett.}{94}{2005}{110403}.

\bibitem{PhysRevLett.95.240404}
\Name{Rizzi M., Rossini D., De~Chiara G., Montangero S. \and Fazio R.}
  \REVIEW{Phys. Rev. Lett.}{95}{2005}{240404}.

\bibitem{PhysRevA.74.035601}
\Name{Apaja V. \and Sylju\aa{}sen O.~F.} \REVIEW{Phys. Rev.
  A}{74}{2006}{035601}.

\bibitem{PhysRevB.77.014503}
\Name{Pai R.~V., Sheshadri K. \and Pandit R.} \REVIEW{Phys. Rev.
  B}{77}{2008}{014503}.

\bibitem{PhysRevLett.102.140402}
\Name{Batrouni G.~G., Rousseau V.~G. \and Scalettar R.~T.} \REVIEW{Phys. Rev.
  Lett.}{102}{2009}{140402}.

\bibitem{PhysRevLett.81.742}
\Name{Ho T.-L.} \REVIEW{Phys. Rev. Lett.}{81}{1998}{742}.

\bibitem{JPSJ.67.1822}
\Name{Ohmi T. \and Machida K.} \REVIEW{J. Phys. Soc. Jpn.}{67}{1998}{1822}.

\bibitem{stenger1998spin}
\Name{Stenger J., Inouye S., Stamper-Kurn D., Miesner H.-J., Chikkatur A. \and
  Ketterle W.} \REVIEW{Nature}{396}{1998}{345}.

\bibitem{Goldenfeld.1992}
\Name{Goldenfeld N.} \Book{Lectures on phase transitions and the
  renormalization group} (Addison-Wesley) 1992.

\bibitem{PhysRevLett.98.160404}
\Name{Lamacraft A.} \REVIEW{Phys. Rev. Lett.}{98}{2007}{160404}.

\bibitem{PhysRevLett.99.120407}
\Name{Uhlmann M., Sch\"utzhold R. \and Fischer U.~R.} \REVIEW{Phys. Rev.
  Lett.}{99}{2007}{120407}.

\bibitem{PhysRevLett.99.130402}
\Name{Damski B. \and Zurek W.~H.} \REVIEW{Phys. Rev. Lett.}{99}{2007}{130402}.

\bibitem{PhysRevA.76.043613}
\Name{Saito H., Kawaguchi Y. \and Ueda M.} \REVIEW{Phys. Rev.
  A}{76}{2007}{043613}.

\bibitem{damski2009quantum}
\Name{Damski B. \and Zurek W.~H.} \REVIEW{New Journal of
  Physics}{11}{2009}{063014}.

\bibitem{0953-8984-25-40-404212}
\Name{Saito H., Kawaguchi Y. \and Ueda M.} \REVIEW{Journal of Physics:
  Condensed Matter}{25}{2013}{404212}.

\bibitem{kibble1976topology}
\Name{Kibble T.~W.} \REVIEW{Journal of Physics A: Mathematical and
  General}{9}{1976}{1387}.

\bibitem{zurek1985cosmological}
\Name{Zurek W.} \REVIEW{Nature}{317}{1985}{505}.

\bibitem{pitaevskii2003bose}
\Name{Pitaevskii L.~P. \and Stringari S.} \Book{Bose-Einstein Condensation}
  (Oxford University Press) 2003.

\bibitem{PhysRevA.75.013607}
\Name{Murata K., Saito H. \and Ueda M.} \REVIEW{Phys. Rev.
  A}{75}{2007}{013607}.

\bibitem{PhysRevLett.92.140403}
\Name{Chang M.-S., Hamley C.~D., Barrett M.~D., Sauer J.~A., Fortier K.~M.,
  Zhang W., You L. \and Chapman M.~S.} \REVIEW{Phys. Rev.
  Lett.}{92}{2004}{140403}.

\bibitem{chang2005coherent}
\Name{Chang M.-S., Qin Q., Zhang W., You L. \and Chapman M.~S.} \REVIEW{Nature
  Physics}{1}{2005}{111}.

\bibitem{PhysRevA.84.063625}
\Name{Guzman J., Jo G.-B., Wenz A.~N., Murch K.~W., Thomas C.~K. \and
  Stamper-Kurn D.~M.} \REVIEW{Phys. Rev. A}{84}{2011}{063625}.

\bibitem{zhou2003quantum}
\Name{Zhou F.} \REVIEW{International Journal of Modern Physics
  B}{17}{2003}{2643}.

\bibitem{PhysRevA.88.043629}
\Name{Phuc N.~T., Kawaguchi Y. \and Ueda M.} \REVIEW{Phys. Rev.
  A}{88}{2013}{043629}.

\bibitem{PhysRevB.40.546}
\Name{Fisher M. P.~A., Weichman P.~B., Grinstein G. \and Fisher D.~S.}
  \REVIEW{Phys. Rev. B}{40}{1989}{546}.

\bibitem{peskin1995introduction}
\Name{Peskin M.~E. \and Schroeder D.~V.} \Book{An introduction to quantum field
  theory} (Westview) 1995.

\bibitem{RevModPhys.76.599}
\Name{Andersen J.~O.} \REVIEW{Rev. Mod. Phys.}{76}{2004}{599}.

\bibitem{beliaev}
\Name{Beliaev S.~T.} \REVIEW{Sov. Phys. JETP}{2}{1958}{299}.

\bibitem{PhysRev.105.1119}
\Name{Lee T.~D. \and Yang C.~N.} \REVIEW{Phys. Rev.}{105}{1957}{1119}.

\bibitem{PhysRev.106.1135}
\Name{Lee T.~D., Huang K. \and Yang C.~N.} \REVIEW{Phys.
  Rev.}{106}{1957}{1135}.

\bibitem{PhysRevA.81.063632}
\Name{Uchino S., Kobayashi M. \and Ueda M.} \REVIEW{Phys. Rev.
  A}{81}{2010}{063632}.

\bibitem{zphys.85.180}
\Name{Hellmann H.} \REVIEW{Z. Phys.}{85}{1933}{180}.

\bibitem{PhysRev.56.340}
\Name{Feynman R.~P.} \REVIEW{Phys. Rev.}{56}{1939}{340}.

\bibitem{PhysRevA.63.053601}
\Name{van Oosten D., van~der Straten P. \and Stoof H. T.~C.} \REVIEW{Phys. Rev.
  A}{63}{2001}{053601}.

\bibitem{rey2003bogoliubov}
\Name{Rey A.~M., Burnett K., Roth R., Edwards M., Williams C.~J. \and Clark
  C.~W.} \REVIEW{Journal of Physics B: Atomic, Molecular and Optical
  Physics}{36}{2003}{825}.

\bibitem{PhysRevLett.96.180405}
\Name{Xu K., Liu Y., Miller D.~E., Chin J.~K., Setiawan W. \and Ketterle W.}
  \REVIEW{Phys. Rev. Lett.}{96}{2006}{180405}.

\bibitem{PhysRevA.77.033628}
\Name{Song J.~L. \and Zhou F.} \REVIEW{Phys. Rev. A}{77}{2008}{033628}.

\end{thebibliography}

\end{document}